\documentclass[prb,twocolumn,aps,amssymb,amsmath,superscriptaddress,showpacs]{revtex4-1}

\usepackage{bm}
\usepackage{graphicx}
\usepackage{color}
\usepackage{mathtools}
\usepackage{amsfonts}
\usepackage{textcomp} 
\usepackage{microtype} 
\usepackage{epstopdf}
\usepackage{hyperref}
\hypersetup{colorlinks=true}
\usepackage[all]{hypcap} 

\begin{document}

\title{Topological Quantum Phase Transition from Fermionic Integer Quantum Hall Phase to Bosonic Fractional Quantum Hall Phase through P-Wave Feshbach Resonance }

\author{Shiuan-Fan Liou}
\affiliation{National High Magnetic Field Laboratory, Florida State University, Tallahassee, FL 32306, USA}
\author{Zi-Xiang Hu}
\affiliation{Department of Physics, Chongqing University,Chongqing, 401331, P. R. China}
\author{Kun Yang}
\affiliation{National High Magnetic Field Laboratory, Florida State University, Tallahassee, FL 32306, USA}

\begin{abstract}
We use exact diagonalization  to study the quantum phases and phase transitions when a single species of fermionic atoms at Landau level filling factor $\nu_f = 1$ in
a
rotating trap interact
through a p-wave Feshbach resonance. We show that under weak pairing interaction, the system undergoes a second order quantum phase transition from $\nu _{f} = 1$
fermionic integer quantum Hall (FIQH) state at positive detuning, to $\nu _{b} = \frac{1}{4}$ bosonic fractional quantum Hall (BFQH) state at negative detuning.
However, when
the pairing interaction increases, a new
phase between them emerges, corresponding to a fraction of fermionic atoms stay in a coherent superposition of bosonic molecule state and an unbound pair. The phase
transition
from FIQH phase to the new phase is of second order
and that from the new phase to BFQH phase is of first order.

\end{abstract}

\date{\today}

\maketitle

{\it Introduction.} Topological phases of matter are of tremendous current interest. Equally important, but perhaps less studied thus far (especially on the experimental
side),
are quantum phase transitions (QPTs) between different topological phases. This is because tuning the system through such QPTs requires great control of certain knobs,
which are
not always available in electronic condensed matter systems. Recently there has been intense activities in topological phases realized in trapped cold atom
systems\cite{NatPhys3, 2010arXiv1007.2677G, PhysRevLett.107.255301,RevModPhys.83.1523, NatPhys12}. Advantages of these systems include availability of qualitatively
different
types of inter-particle interactions, and great abilities experimentalists have to control them. This allows for detailed studies of QPTs in such systems. Among such
activities
are attempts to realize quantum Hall (QH) states\citep{Avdphys57.539, Scientific1}, which were the very first topological phases discovered.

In electronic systems QPTs between different QH phases are usually driven by changing magnetic field, and disorder plays a dominant role at these transitions. In fact
the
field-driven transition can often be mapped to a transition at fixed field but driven by disorder strength instead. The presence of disorder, especially when combined
with
electron-electron interaction, makes such transitions extremely hard to study theoretically. Over the years theorists instead studied possible QH transitions driven by
interactions {\em without} disorder\cite{PhysRevLett.70.1501, PhysRevB.48.13749, PhysRevLett.84.3950}; in most cases these require an additional periodic potential that
is not
available in electronic systems.

It was pointed out by Yang and co-workers\cite{PhysRevLett.100.030404, PhysRevLett.106.170403}, that in cold atom systems one can realize some of the closely related QH
transitions with {\em neither} disorder {\em nor} periodic potential. Instead what drives the transitions are Feshbach resonances, across which two fermions bind into a
bosonic
molecule. The two phases involved are a
fermionic integer QH phase on the atomic side (positive detuning), and bosonic fractional QH phase on the molecular side (negative detuning). One such
phase transition\cite{PhysRevLett.100.030404} driven by s-wave Feshbach resonance has been confirmed numberically\cite{arXiv:1612.09184}. The other transition from $\nu
_{f} = 1$ fermionic integer quantum Hall (FIQH) phase to $\nu _{b} = \frac{1}{4}$ bosonic fractional quantum Hall (BFQH) was proposed by Barlas and
Yang\cite{PhysRevLett.106.170403}, where they studied this phase transition using quantum field theory methods, and argued that it is a second-order QPT in the
(2+1)-dimensional
Ising universality class.

In this paper we use exact diagonalization (ED) method to study the model of Barlas and Yang\cite{PhysRevLett.106.170403}. The purpose of such numerical study is
three-fold.
First of all, while field theory approaches may determine the possible phases, it cannot determine the actual location of the phase boundary; the latter needs to be
determined
by  calculations based on microscopic models. Secondly while field theory can tell if a second-order phase transition is possible, in general it does not guarantee it
must
happen in a specific microscopic model; in fact first order transitions are {\em always} possible. Lastly since field theory only keeps certain
long-wavelength/low-energy
degrees of freedom, it might miss certain phases. Indeed through our numerical work we (i) determine the phase diagram quantitatively; (ii) confirm that the direct
transition
from FIQH phase to BFQH phase is 2nd order; and (iii) perhaps most importantly, find an intermediate phase between FIQH and BFQH phases in certain region of the phase
diagram.

To investigate this QPT, we consider a single species of fermions confined to two-dimensions and  under either rotation or synthetic gauge field, whose effects mimic that
of a
strong magnetic field. We focus on the case where the Landau level filling factor $\nu _{f} = 1$. When the system is tuned through p-wave FR, two fermions can pair up and
form a
p-wave bosonic molecule with twice the ``charge". We assume the Landau level spacing is large enough such that we can assume all particles are confined to their
respective
lowest Landau level.
The system can thus be described by the following Hamiltonian on a disc (which is the geometry of our choice for finite-size numerical study, where total angular momentum
is a
good quantum number due to rotational symmetry):\\
\begin{equation}\label{eq_H}
\begin{aligned}
H &= \delta \sum\limits _{m} (b _{m} ^{\dagger} b _{m} - 2 f _{m} ^{\dagger} f _{m}) \\
   & + \left( \, \sum\limits _{m _{1}, m _{2}, m _{3}} g _{m _{1}, m _{2}, m _{3}} b _{m _{1}} ^{\dagger} f _{m _{2}} f _{m _{3}} + h.c. \, \right) \\
   & + \sum\limits _{m _{1}, m _{2}, m _{3}, m _{4}} v ^{(0)} _{m _{1}, m_{2}, m _{3}, m _{4}} \, b _{m _{1}} ^{\dagger} b _{m _{2}} ^{\dagger} b _{m _{3}} b _{m _{4}}
   \\
   & + \sum\limits _{m _{1}, m _{2}, m _{3}, m _{4}} v ^{(2)} _{m _{1}, m_{2}, m _{3}, m _{4}} \, b _{m _{1}} ^{\dagger} b _{m _{2}} ^{\dagger} b _{m _{3}} b _{m _{4}}
\end{aligned}
\end{equation}
where $\left\lbrace m _{i}\right\rbrace$ are angular momentum quantum numbers of the lowest Landau level single particle orbitals, with $b _{m}$ ($b ^{\dagger} _{m}$) and
$f
_{m}$ ($f ^{\dagger} _{m}$) the corresponding annihilation (creation) operators for bosons and fermions. The first term above is the chemical potential term in which
$\delta$ is
the detuning corresponding to the energy difference between a pair of fermions and a boson; this term determines whether the atoms should pair up to form molecules or
stay
unbound energetically.

The second term depicts the pairing interaction through p-wave FR. The matrix element $g _{m _{1}, m _{2}, m _{3}} = g \, \delta _{m _{1}, M} \, \left\langle 1, M| m
_{2}, m
_{3} \right\rangle$, where $g$ represents the strength of the pairing interaction in the p-wave channel (corresponding to width of the p-wave FR), $|m _{2}, m _{3}
\left.\right\rangle$ is a two-body state with the two particles having angular momentum $m_2$ and $m_3$ respectively, and
$|1, M \left.\right\rangle$ is a two-body state with their relative angular momentum equal to 1 and center-of-mass angular momentum $M$. This matrix element indicates
that only
those pairs of fermions whose relative angular momentum $\Delta m = 1$ can pair up, and the consequently composite boson has orbital angular momentum $m_1=M= m _{2} + m
_{3} -
1$. This term conserves angular momentum because the p-wave bosonic molecule also carries an internal momentum $+1$.
The matrix element $\left\langle 1, M| m _{2}, m _{3} \right\rangle$ is a special case of

\begin{equation}
\begin{aligned}
&\left\langle \Delta m , M | m _{1}, m _{2} \right\rangle \equiv \\
&\frac{1}{\sqrt{(2 \pi) ^{4} \cdot 2 ^{\left( \Delta m + M + m _{1} + m _{2} \right)} \cdot \Delta m! \cdot M! \cdot m _{1}! \cdot m _{2}! \cdot l ^{8}}} \\
&\int d ^{2} z_{1} \int d ^{2} z _{2} \left( \frac{z _{1} ^{\ast} - z _{2} ^{\ast}}{\sqrt{2} l} \right) ^{\Delta m} \left(\frac{z _{1} ^{\ast} + z _{2} ^{\ast} }{\sqrt{2}
l}
\right) ^{M} \left( \frac{z _{1}}{l} \right) ^{m _{1}} \left( \frac{z _{2}}{l} \right) ^{m _{2}} \\
& \hspace*{6cm} \times e^{- \frac{|z _{1}| ^{2} + |z _{2}| ^{2}}{2 l ^{2}}}  \\
& = \, \sqrt{\frac{\Delta m! \, M!}{\left( 2 \pi \right) ^{4} \cdot 2 ^{\Delta m + M} \, \cdot m _{1}! \cdot m _{2}!}} \\
& \sum\limits_{n} \, (-1) ^{\Delta m - n} \, C ^{m _{1}} _{n} \, C ^{m _{2}} _{\Delta m - n} \,  \delta _{\Delta m + M, m _{1} + m _{2}}
\end{aligned}
\end{equation}
where $z _{a} \equiv x _{a} + i y _{a}$, is the complex coordinate of the $a$-th particle on a disc in the lowest Landau level, $d ^{2} z _{a} = dx _{a} dy _{a}$, $l$ is
the
corresponding magnetic length, and $\Delta m$ is the relative angular momentim of a pair. The sum of $n$ is summing over all natural numbers bounded at max$\left(0,
\Delta m - m
_{2} \right) \leq n \leq$ min$\left(\Delta m, m _{1} \right)$. Note that the bosonic magnetic length square $l _{b} ^{2}$ is half of the fermionic magnetic length square
$l _{f}
^{2}$ ($l _{b} ^{2} = l _{f} ^{2} /2$) in our model due to the doubled charge of bosons.

The last two terms correspond to {\em repulsive} interactions between bosons. In order to stabilize the BFQH state with $\nu _{b} = \frac{1}{4}$, we consider the zero-th
order
and the second order of Haldane pseudo-potentials~\cite{PhysRevLett.51.605}. The matrix element, $v _{m _{1}, m_{2}, m _{3}, m _{4}} ^{(\alpha)} = v ^{(\alpha)}
\sum\limits _{M}
\left\langle m _{1}, m _{2}|\alpha, M \right\rangle \left\langle \alpha, M| m _{3}, m _{4} \right\rangle$, where $v ^{(\alpha)}$ is the strength of the $\alpha$-th order
Haldane
pseudo-potential. In our calculation, we simply use $v ^{(0)} = v ^{(2)} = 1$.

In this model, the total charge ($N _{tot}$) and the total angular momentum ($M _{tot}$) are good quantum numbers. The total charge is the sum of the number of fermions
and
twice the number of bosons:
\begin{equation}
N _{tot} \, = 2N_b + N_f=\, \sum\limits_{m} \, (2 b _{m} ^{\dagger} b _{m} + f _{m} ^{\dagger} f _{m}).  \\
\end{equation}
The prefactor 2 in the bosonic part comes from the fact that each bosonic molecule has twice the charge of the fermion. And the total angular momentum is the sum over
all
angular momentum of orbitals occupied by bosons and fermions plus the number of bosons since each boson has one internal angular momentum:
\begin{equation}
M _{tot} \, = \, \sum\limits_{m} \, \left[ \, (m + 1) b _{m} ^{\dagger} b _{m} \, + m f _{m} ^{\dagger} f _{m} \, \right]. \\
\end{equation}
In our numerical calculation, we use these two quantum numbers to label the sector in which we perform our calculations.

There are two limits of this Hamiltonian in Eq.~(\ref{eq_H}). For $\delta > 0$ and $|\delta| \gg g$, a boson costs more energy than an unbound pair of fermions, so the
ground
state will be dominated by fermions and the system forms a FIQH state at $\nu_f=1$, which does not require interactions between fermions to stabilize (and that is the
reason we
do not include fermion interaction in our model).
On the other hand, when $\delta <0$ and $|\delta| \gg g$ it is energetically favorable for fermions to pair into bosonic molecules. The resultant boson filling factor is
$\nu_b=1/4$, resulting in a Laughlin-type BFQH state with the boson-boson interaction we introduced.
We can easily distinguish between these two phases by inspecting their low-energy spectra, as we now turn to.

\begin{figure}[!ht]
\centerline{\includegraphics[width=8cm]{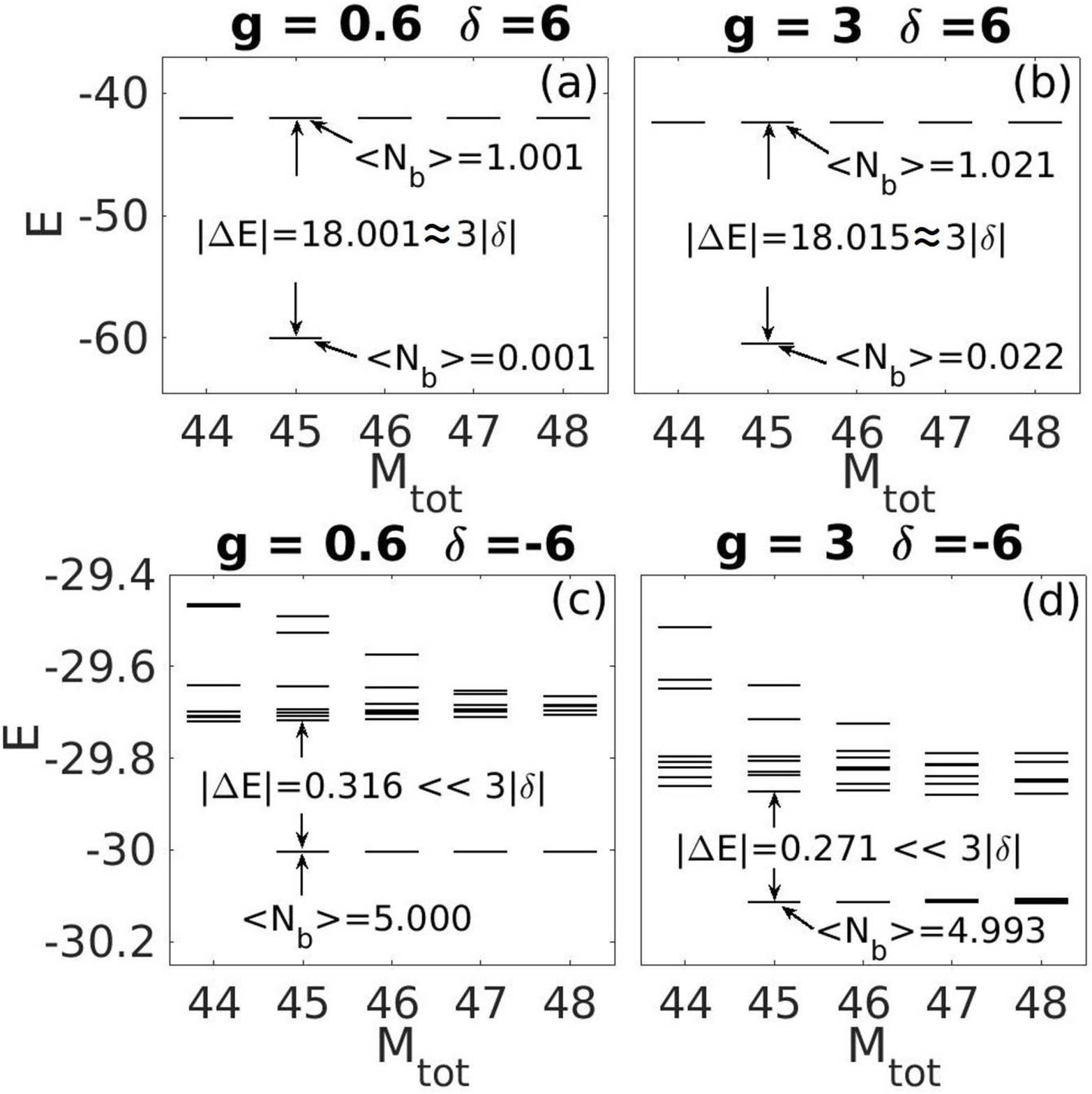}}
\caption{The energy spectra for a system with $N_{tot}=10$ fermions, given 10 fermionic orbitals and 20 bosonic orbitals. 8 lowest energy states are plotted for each $M
_{tot}$.
The ground state at $M_{tot}=M_{gs}=45$ is separated by a large gap from all excited states for (a) $g$ = 0.6 and $\delta = 6$ and (b) $g$ = 3 and $\delta = 6$, which are
under
FIQH limit. We find a set of low-lying states for $M _{tot}\ge M_{gs}=45$ for (c) $g$ = 0.6 and $\delta = -6$ and (d) $g$ = 3 and $\delta = -6$, which are under BFQH
limit.
These correspond to the Laughlin-like state at $M_{tot}=M_{gs}=45$ and edge states for $M _{tot} > M_{gs}=45$. The number of these states indicated in the plots match
the
expected number of edge states. $\left\langle N _{b} \right\rangle$s indicate the expectation values of boson numbers in the corresponding states pointed by small
arrows.}
\label{eng_spctm}
\end{figure}

In Fig.~\ref{eng_spctm}, we show the energy spectra for a system with $N _{tot} = 10$ given 10 fermionic orbitals and 20 bosonic orbitals with $M _{tot}$ from 44 to 48.
Under
FIQH limit, the system will have the lowest energy state only when it forms a FIQH state in which there are no bosons and all fermionic orbitals are occupied, namely $M
_{tot} =
M _{gs}$ with \\
\begin{equation}
M _{gs} = \frac{N _{tot} \left( N _{tot} - 1 \right)}{2}.
\end{equation}
$M _{gs}$ is 45 in this case for $N_{tot} = 10$. Because we only give the least number of fermionic orbitals which is 10 here for FIQH state, no edge states should show
up at $M
_{tot} > M _{gs}$. In Fig.~\ref{eng_spctm} (a) and (b) with $\delta = 6$ (FIQH limit) at $g = 0.6$ (weak pairing) and $g$ = 3 (strong pairing), the lowest energy states
indeed
appear at $M _{tot} = 45$ and the expectation values of boson numbers in the lowest energy states, $\left\langle N _{b} \right\rangle$, are close to zero as well.
Furthermore,
the first excited state is a state with one boson. The energy difference between the ground state and the lowest energy excited states, mainly due to the chemical
potential
term, is about 3$\delta$ (losing two fermions and gaining one boson) as found in Fig.~\ref{eng_spctm} (a) and (b).

In BFQH limit, if the system which has $N _{b}$ bosons forms a BFQH state with $\nu = \frac{1}{4}$, it will have Laughlin wave function as
\begin{equation}
\psi _{\nu = \frac{1}{4}} \left( z _{1}, z _{2} \cdots \right) \, = \, \left[ \prod _{i < j} ^{N_b} \, \left( z _{i} - z _{j} \right) ^{4} \right] \, e ^{-\sum\limits _{k
= 1}
^{N} \, \frac{|z _{k}| ^{2}}{4 \, l ^{2}}}.
\end{equation}
The total angular momentum is composed of two parts: the angular momentum of the orbitals occupied by bosons and the internal angular momentum of bosons, $M _{tot} = 4
\times N
_{b} \left( N _{b} - 1 \right) / 2 + N _{b}$. In this limit, $N _{b} = N _{tot} / 2$, so $M _{tot} = M _{gs}$, the same as that in FIQH state. When $M _{tot} < M _{gs}$,
there
is no state with lower energy than the Laughlin state at $M _{gs}$. However, when $M _{tot} > M _{gs}$, edge states~\cite{wen:IJMPB} degenerate with the Laughlin state
exist.
Therefore, by counting and comparing the numbers of the low-lying states with the numbers of edge states of Laughlin type states at various $M _{tot}$, we can demonstrate
the
system forms a BFQH state. In Fig.~\ref{eng_spctm} (c) and (d) with $\delta = -6$ at $g = 0.6$ and $g = 3$ we demonstrate this is indeed the case. Notice that the
low-lying
states are no longer exactly degenerate because of the existence of the pairing interaction. Besides, the boson number in the low-lying state at $M _{tot} = 45$ is very
close to
5, the maximum number of bosons, as expected. In Fig.~\ref{eng_spctm} (d), we see that $\left\langle N _{b} \right\rangle$ has bigger deviation from 5. This is due to
the
fluctuation induced by the strong pairing interaction namely fermions have bigger matrix element to go back and forth between paired and unpaired states. This is also the
reason
why the average boson number is bigger in the FIQH state with larger $g$, as shown in Fig. ~\ref{eng_spctm} (b).
The other thing to notice is in the BFQH regime the low-energy excited state still have essentially all particles as bosons,
as a result their excitation energies in Fig.~\ref{eng_spctm} (c) and (d) are much smaller than $3|\delta|$ and determined by the boson-boson interaction instead.

\begin{figure}[!ht]
\centerline{\includegraphics[width=8.5cm]{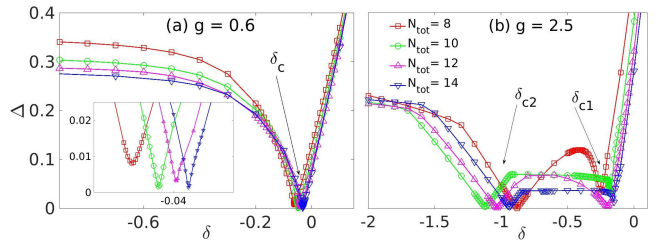}}
\caption{$\left(\right.$ Color online $\left.\right)$ Plot of the energy gap ($\Delta \equiv E _{1} - E _{0}$, where $E _{0}$ and $E _{1}$ are the energies of the ground
state
and the first excited state) versus $\delta$ for systems with $N _{tot} = 8, 10, 12$ and 14 at $M _{gs}$ for (a) $g$ = 0.6. There is one gap closing point for each curve.
Inset:
blow-up of the gap-closing region. (b) For $g$ = 2.5. There
are two gap closing points for each curve.}
\label{vars_size}
\end{figure}

\begin{figure}[!ht]
\centerline{\includegraphics[width=8cm]{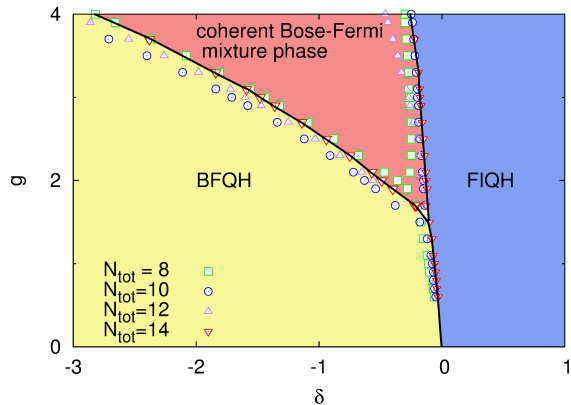}}
\caption{$\left(\right.$ Color online $\left.\right)$ Phase diagram for systems with $N _{tot} = 8, 10, 12$ and 14. At small g, in the region $\delta > 0$ is FIQH phase;
in the
region $\delta < 0$ is BFQH phase. At large g, the coherent Bose-Fermi mixture phase appears between FIQH and BFQH phases.}
\label{ph_dm}
\end{figure}

In our model for a specific $N _{tot}$, we need at least $N _{tot}$ fermionic orbitals and $2N _{tot} - 3$ bosonic orbitals respectively in order to access the
appropriate
ground states in the FIQH and BFQH limits. Furthermore, $M _{gs}$ is the same in these two limits. Therefore we will focus on the $M _{gs}$ sector in our calculations
from now
on using the minimum orbital numbers mentioned above, unless noted otherwise.
To identify phase boundaries, we drive a system from FIQH phase to BFQH phase by changing $\delta$ at various $g$.
Since quantum phase transitions between different gapped phases must involve gap closing, we plot the gap $\Delta$ of systems in Fig.~\ref{vars_size} (a) and (b) as
functions of
$\delta$, where the gap is defined as the energy
difference between the first excited state and the ground state in the $M _{gs}$ sector.
Four system sizes with $N_{tot}= 8, 10, 12, 14$ are studied. There is one gap-closing point at $g$ = 0.6 (weak pairing regime, Fig.~\ref{vars_size} (a)), indicating a
single
phase boundary separating the FIQH phase and the BFQH phase. Closer inspection (see inset of Fig.~\ref{vars_size} (a)) the gap is not exactly zero in these finite size
calculations, but instead approaches zero with increasing systems sizes, indicating the transition is due to level anti-crossing, consistent with a 2nd order QPT
predicted by
Barlas and Yang~\cite{PhysRevLett.106.170403}. Unfortunately the limited system sizes and obvious fluctuations that are present do not allow us to perform finite-size
scaling to
extract critical exponents.

The situation is quite different at $g$ = 2.5 (strong pairing regime). In this case
the gap closes twice, implying that the systems undergo two phase transitions in the strong pairing regime.
We provide a phase diagram in Fig.~\ref{ph_dm}, based on the locations of gap-closing points.
As predicted by Barlas and Yang~\cite{PhysRevLett.106.170403}, there is a direct transition
from FIQH state to BIQH state in the weak pairing regime (small $g$).
However, it is clear that an unexpected new phase shows up between FIQH phase and BFQH phase in the strong pairing regime (large $g$). We now study the properties of this
new
phase, and phase transitions involving it.

\begin{figure}[!h]
\centerline{\includegraphics[width=9cm]{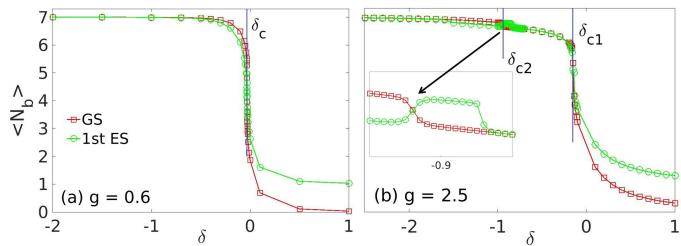}}
\caption{$\left(\right.$ Color online $\left.\right)$ The expectation values of bosons, $\langle N _{b} \rangle$, in the ground state and the first excited state versus
$\delta$
for the system with $N _{tot} = 14$ at (a) $g$ = 0.6 and (b) $g$ = 2.5. The vertical blue lines locate the critical points.}
\label{fig4}
\end{figure}

\begin{figure}[!h]
\centerline{\includegraphics[width=7cm]{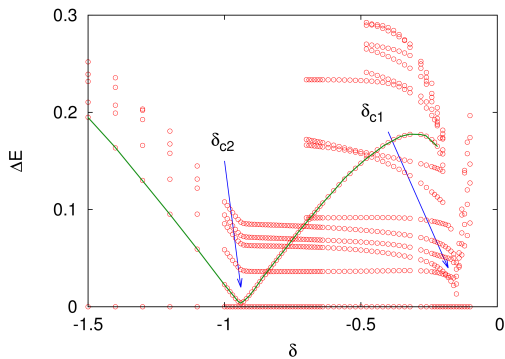}}
\caption{$\left(\right.$ Color online $\left.\right)$ Energy spectra for the system with $N _{tot} = 14$ at $g$ = 2.5. $\Delta E = E _{i} - E _{0}$ where $E _{i}$ denotes
the
energy of the $i$-th state and $E _{0}$ is the
ground state energy.}
\label{gap_nb_snf_14_2p5_all}
\end{figure}

As discussed earlier, one can distinguish between FIQH and BFQH states by the average number of bosonic molecules in the ground state $\langle N_b \rangle$. To gain
intuition
into the new phase, we inspect $\langle N_b \rangle$ for both the ground and first excited states. Fig.~\ref{fig4} (a) shows that for $N _{tot} = 14$ at $g$ = 0.6 (weak
pairing
regime), $\langle N_b \rangle$s in the ground state and the first
excited state increase monotonically and smoothly from FIQH state toward BFQH state, consistent with a second order phase transition happening at the critical point,
$\delta
_{c}$. In the strong pairing regime in Fig.~\ref{fig4} (b), $\langle N_b \rangle$ still changes smoothly at $\delta _{c1}$.
However, $\langle N_b \rangle$s in the ground state and
the first excited state have jumps at $\delta _{c2}$ with opposite signs.
This suggests a first-order phase transition triggered by crossing between the ground and first excited states at $\delta _{c2}$. Motivated by this we examine the energy
spectrum for $N _{tot} = 14$ case at
$g$ = 2.5 shown in Fig.~\ref{gap_nb_snf_14_2p5_all}. It clearly shows that a high energy state comes down and eventually becomes the ground state after a sequence of
level
crossings with other lower energy states as $\delta$ varies from $\delta _{c1}$ toward $\delta _{c2}$. This suggests
the phase transition occurring at $\delta _{c2}$ is a first order phase transition, triggered by crossing of the Laughlin-like state with the ground state of the new
phase
between the BFQH state and FIQH state.

Returning to the boson number expectation value, we expect the ground state in the intermediate phase contains a finite fraction of fermions that stay in a coherent
superposition of bound molecular state and unbound scattering state, even when $\delta$ is quite negative. Such a superposition state gains energy from the pairing term,
and
when $g$ is much larger than $v ^{(0)}$ and $v ^{(2)}$, the pairing term dominates, and the energy gain from such coherence dominates the energy cost from putting
fermions in
the unpaired state. The physics of this intermediate phase is perhaps most clearly revealed by considering the limit $g\rightarrow\infty$. In this limit there is no way
to tell
whether a fermionic atom is in an (open channel, unbound) scattering state, or forms a bound (closed channel, molecular) state with another fermion, because the $g$-term
in the
Hamiltonian forces all p-wave fermion pairs to be in a coherent superposition of these two states. For this reason we call it coherent Bose-Fermi mixture phase. A
natural
variational wave function for the ground state with total angular momentum $M _{tot} = \sum\limits_{\left\lbrace M \right\rbrace} (M+1)$ is
\begin{equation}
\begin{aligned}
\psi &= \, \prod _{\left\lbrace M \right\rbrace} \, \left( u_M b ^{\dagger} _{M} + v_M \sum\limits_{m _{1}, m _{2}} \, \left\langle m _{1}, m _{2} | 1, M \right\rangle \,
f
^{\dagger} _{m _{1}} f ^{\dagger} _{m _{2}} \right) |0 \left. \right\rangle\\
\end{aligned}
\end{equation}
where $\left\lbrace M \right\rbrace$ is a set of boson angular momenta, with $u_M$ and $v_M$ variational parameters that describe the coherence between boson and fermion
pair
states. Since there exist many different configurations of $\left\lbrace M \right\rbrace$, the system is most likely compressible. Of course the actual ground state in
the $M
_{tot}$ sector can be written as linear superpositions of the state above with different configurations of $\left\lbrace M \right\rbrace$.~\cite{PhysRevB.79.224404}

\begin{figure}[!ht]
\centerline{\includegraphics[width=8.7cm]{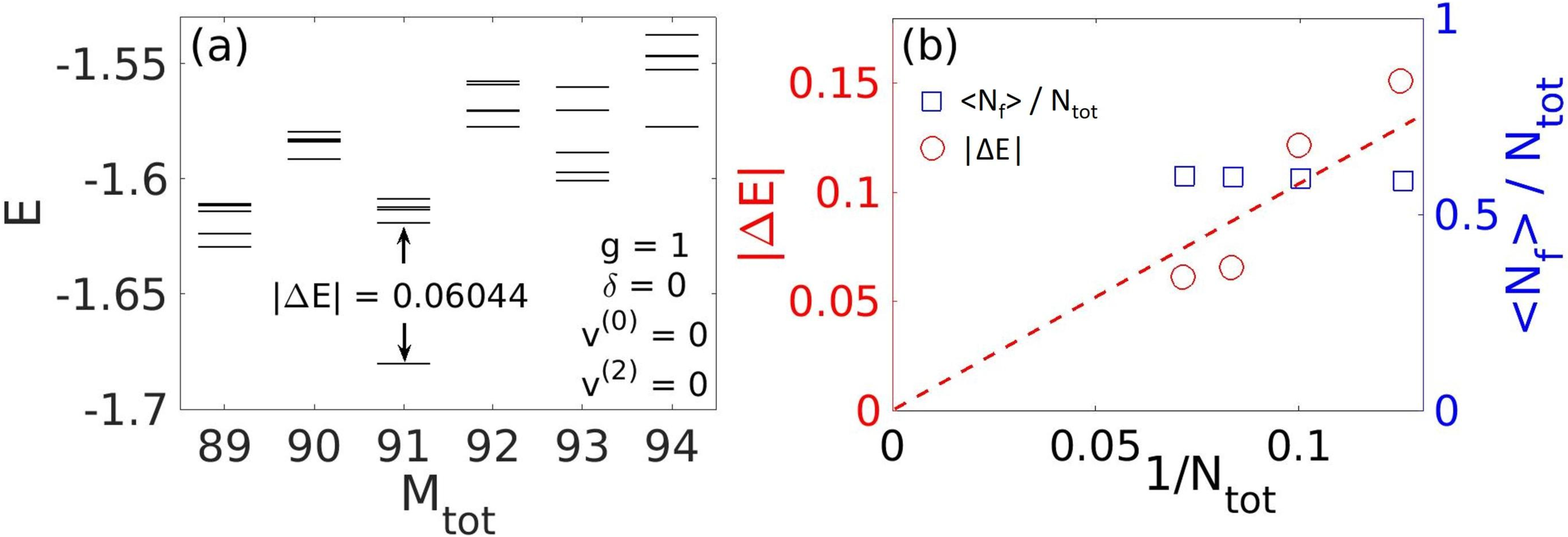}}
\caption{(a)$\left(\right.$ Color online $\left.\right)$ Energy spectra for $N_{tot} = 14$ with 14 fermionic orbitlas and 25 bosonic orbitals which are the least orbital
numbers
and parameters $g = 1$, $\delta = 0$ and $v ^{(0)} = v ^{(2)} = 0$. (b) Energy gap between the lowest energy state and the first excited state $|\Delta E|$ versus system
size on
the left axis and average fermion fraction $\frac{\left\langle N_{f} \right\rangle}{N _{tot}}$ versus system size on the right axis. The black dashed line extrapolates
the
energy gap to origin. The fermion fraction is independent of system size.}
\label{new_phase}
\end{figure}

To test the analysis above, we examine the energy spectrum for $N_{tot} = 14$ of an extreme case with $g = 1$ and all $\delta$ and $v ^{(0)}$ and $v ^{(2)}$ equal to 0
at
various $M _{tot}$ in Fig.~\ref{new_phase}(a). We see the lowest energy state locates at $M _{gs}$ which is 91 in this case, just like the neighboring BFQH and FIQH
phases.
The fermion fraction of the total conserved charge is about 0.6 and independent of the system size, as shown in Fig.~\ref{new_phase}(b). This supports the argument that
the
$g$-term enforces the ground state to be a superposition of a scattering state and a bound state with $u_{M} \approx v_{M}$.
On the other hand the gap $|\Delta E|$ decreases significantly as system size $N_{tot}$ increases; it plausibly extrapolates to zero upon approaching thermodynamic limit.
This
strongly suggests that the coherent Bose-Fermi mixture phase is compressible.

\vspace*{0.3cm}

{\it Conclusion.}
We performed a systematic numerical study of the topological phase transition from an integer quantum Hall (FIQH) state made of fermionic atoms, to a bosonic fractional
quantum
Hall (BFQH) state made of bosonic molecules, driven by a p-wave Feshbach resonance.
The phase diagram can be separated into two regimes: weak pairing and strong pairing. In the weak pairing
regime corresponding to narrow resonance, we demonstrate the existence of the second order quantum phase transition from FIQH phase to BFQH phase
which is consistent to earlier theoretical work~\cite{PhysRevLett.106.170403}. In the strong
pairing regime corresponding to wide resonance, a new phase appears which contains
a finite fraction of fermions that stay in a coherent super-position of bound molecular state and unbound scattering state.
According to the energy spectra and the behavior of $\langle N_b \rangle$, we conclude that the phase
transition from FIQH state to the new phase is of second order; from the new phase to BFQH phase it is a first order transition due to the
energy level crossing behavior.

{\it Acknowledgments.} S.-F. Liou and K. Yang are supported by DOE grant No. de-sc0002140. Z-X. Hu is supported by NSFC under Project No.
11674041, 91630205 and FRF for the Central Universities No. CQDXWL-2014-Z006.

\bibliography{ref}

\end{document}